\documentclass[5p,times]{elsarticle}
\usepackage[T1]{fontenc}
\usepackage[utf8]{inputenc}
\usepackage{microtype}
\usepackage{amsmath,amsfonts,amssymb}
\usepackage{graphicx}
\usepackage{booktabs}
\usepackage{multirow}
\usepackage{makecell}
\usepackage{array}
\usepackage{xcolor}
\usepackage{colortbl}
\usepackage{algorithm}
\usepackage{algorithmic}
\usepackage{url}
\usepackage{float}
\usepackage{subcaption}
\usepackage{placeins}
\usepackage{tikz}
\usepackage{lineno}

\usetikzlibrary{positioning, fit, arrows.meta, calc, shapes.geometric,
                 decorations.pathreplacing, backgrounds}

\definecolor{xlsrblue}{HTML}{2C5F8A}
\definecolor{probegold}{HTML}{D4A84B}
\definecolor{classgreen}{HTML}{3A7D44}
\definecolor{fusered}{HTML}{B04040}
\definecolor{lightblue}{HTML}{D6E8F5}
\definecolor{lightgold}{HTML}{F5EDD6}
\definecolor{lightgreen}{HTML}{D6EDDA}
\definecolor{lightred}{HTML}{F5D6D6}
\definecolor{graybox}{HTML}{F0F0F0}
\definecolor{darkgray}{HTML}{4A4A4A}

\newcommand{\balacc}{\text{BalAcc}}

\newcommand{\layerset}{\{6, 7, 17, 19\}}

\begin{document}

\begin{frontmatter}

\title{Probing-Guided Layer Selection from Self-Supervised Speech Models for Generalizable Audio Deepfake Detection}

\author[mtu]{Marjan~Beheshti\corref{cor1}}
\cortext[cor1]{Corresponding author.}
\ead{mbehes2@mtu.edu}
\author[mtu]{Majid~Rostami}
\author[mtu]{Bo~Chen}

\address[mtu]{Department of Computer Science, Michigan Technological University, Houghton, MI, USA}

\begin{abstract}
Audio deepfake detection systems often fail to generalize across domains because they rely on features tied to specific attacks or recording conditions. Self-supervised speech models offer rich multi-layer representations, yet existing approaches either use a single layer or fuse all layers indiscriminately, and only reveal layer importance after training.

We propose a model-agnostic, two-stage methodology that identifies informative depth zones before any task-specific model is trained. In the first stage, lightweight XGBoost probes evaluate each transformer layer's cross-domain discriminative power, producing a layer ranking. In the second stage, a compact neural classifier fuses only the selected layers through per-layer attention pooling and a shared bottleneck projection, while the backbone remains frozen. Applied across three backbones, the probing reveals two key findings. First, informative layers cluster in depth \emph{zones} rather than at uniquely optimal positions: within-zone substitutions fall within multi-seed noise, while zone violations degrade performance by up to 5$\times$. Second, the probing produces backbone-specific selections rather than a fixed layer recipe. On XLS-R-300M, four probing-selected layers with 1.34M trainable parameters achieve 4.94$\pm$0.32\% equal error rate on In-The-Wild and 5.07\% cross-domain average over four shared datasets, a 28\% relative improvement over the best prior frozen-backbone result (Xiao and Vu, 2025) using all 25 layers with identical training data.
\end{abstract}


\end{frontmatter}


\section{Introduction}

Modern text-to-speech and voice conversion systems can clone a speaker's voice from as little as three seconds of audio~\cite{wang2023valle}, producing fakes that human listeners cannot reliably distinguish from real speech~\cite{yi2023survey}. The threat is already real: in early 2024, an employee at the engineering firm Arup authorized \$25 million in wire transfers during a video call where every other participant was a deepfake~\cite{iproov2024}, and industry estimates place deepfake-enabled fraud losses in the hundreds of millions of dollars per year~\cite{deloitte2024,resembleai2025}. Beyond fraud, voice cloning enables threats ranging from CEO impersonation to bypassing biometric authentication~\cite{almutairi2022,masood2023}. Since humans cannot reliably detect these fakes, automated detection is essential. Crucially, such detectors must generalize across unseen attacks and recording conditions.

We define a \emph{domain} as the combination of attack algorithms, recording conditions, and speaker populations that characterize a dataset. \emph{In-domain} evaluation tests on data drawn from the same domain as training, while \emph{cross-domain} evaluation tests on data from entirely different domains---unseen attack types, different acoustic environments, or different speaker populations. %
Most existing deepfake detectors focus on in-domain accuracy, but maintaining detection performance across such domain shifts is still an open problem. 
M\"uller et al.~\cite{muller2022itw} constructed the In-The-Wild dataset from real-world deepfake attacks found online, but their detectors that achieved near-perfect accuracy on ASVspoof benchmarks degraded to error rates greater than 30\% on unseen samples. The degradation appears across architectures. AASIST~\cite{jung2022aasist}, a spectro-temporal graph attention network, achieves 0.83\% equal error rate (EER) on ASVspoof 2019 LA Eval but increases to 34.8\% on In-The-Wild~\cite{yi2023survey}. RawNet2~\cite{tak2021rawnet2} follows a similar degradation trajectory. These models learn to detect specific vocoder artifacts present in their training data rather than acquire a transferable notion of what makes speech synthetic.


This cross-domain collapse stems from the feature extractor. Conventional front-ends, whether handcrafted or learned end-to-end, are built on a narrow acoustic distribution and overfit to the specific vocoder signatures present in the training data. Self-supervised speech models offer a fundamentally different starting point. Pre-trained on hundreds of thousands of hours of diverse, multilingual speech through task-agnostic objectives, models in the wav2vec~2.0 family \footnote{The wav2vec~2.0 family includes Base (12 layers, $d{=}768$) and Large (24 layers, $d{=}1024$) configurations. XLS-R additionally offers 1B (48 layers) and 2B (48 layers) variants. WavLM follows the same Base/Large split. In this work, we use exclusively the Large configurations: XLS-R-300M, WavLM Large, and XLSR-53, all with 24 transformer layers and $d{=}1024$.}~\cite{baevski2020wav2vec}---including XLSR-53, XLS-R~\cite{babu2022xlsr}, and WavLM~\cite{chen2022wavlm}---learn representations that capture general properties of speech rather than dataset-specific artifacts. Crucially, these representations are not monolithic.
Pasad et al.~\cite{pasad2021layerwise,pasad2023comparative} demonstrated that the 24 transformer layers encode qualitatively different information along a  consistent gradient: early layers capture acoustic and spectral properties, middle layers encode phonetic and prosodic structure, and late layers represent higher-level linguistic content. This hierarchy matters for deepfake detection because synthesis artifacts are themselves heterogeneous. Vocoder-based attacks leave spectral traces best captured by early-to-mid layers~\cite{sun2023vocoder}, while TTS systems introduce statistical irregularities in prosody and linguistic structure that only late layers represent~\cite{todisco2019asvspoof}. Consequently, the 24 layers vary widely in their utility for detection, raising the fundamental questions that this paper aims to address: \emph{1) which layers carry the most discriminative information for cross-domain detection, and 2) how should they be combined?}

Not all layers contribute equally: El~Kheir et al.~\cite{elkheir2025layerwise} found that truncating to the first 10--12 layers matches or exceeds full-model performance, and Serrano et al.~\cite{serrano2025layer} showed that all-layer fusion degrades out-of-domain EER compared to the best individual layer. Yet existing work has explored three strategies for layer selection, none of which exploits this observation directly. The \textit{first} strategy selects a single layer, typically the last hidden state or one chosen by validation~\cite{pascu2024,wang2024vocoded}. This is practical and has proven effective on benchmarks where training and test distributions are similar, but restricts the detector to a single level of abstraction, discarding the acoustic, phonetic, or linguistic representations encoded at other depths. The \textit{second} strategy fuses all layers. Mart\'in-Do\~nas et al.~\cite{martindonas2024} learn a weighted combination of all 25 hidden states, and Pan et al.~\cite{pan2024attentive} apply attention-based merging across all layers. Both learn implicit layer weights but include all representations, even those that Xiao and Vu~\cite{xiao2025layerwise} showed individually produce EER above 30\% on In-The-Wild. The \textit{third} strategy, also from Xiao and Vu~\cite{xiao2025layerwise}, trains 25 separate one-class softmax classifiers and sums their output scores at inference. Their post-hoc analysis confirms that layers 4--8 and 18--21 carry most of the discriminative signal, but the architecture still uses all 25 layer classifiers with equal score contribution, rather than removing the layers their own analysis identifies as uninformative. Across all three strategies, layer selection is absent or performed only after the full model is trained, providing no diagnostic information on which depth zones carry cross-domain discriminative information. 

The core difficulty is that layer importance is not uniform across domains: as we show in Section~\ref{sec:probing_results}, no single layer dominates all evaluation datasets, and in-domain validation alone cannot reliably identify which layers will generalize. In this work, we propose a \textit{two-stage} approach that addresses this gap through \textit{principled layer selection} before classifier training. Specifically in the first stage, we train independent lightweight classifiers on the frozen hidden states of each self-supervised learning (SSL) 
transformer layer and evaluate them across multiple cross-domain datasets, producing a layer ranking grounded in empirical discriminative power. From the top-ranked layers, we select a compact subset that spans multiple representational depths. In the second stage, we extract only these four layers' hidden states and process each through attention pooling and a shared bottleneck projection. The resulting embeddings are concatenated and fed into a single neural classifier with 1.34M trainable parameters, while the entire XLS-R backbone remains frozen.

On the In-The-Wild benchmark, this approach achieves 4.94$\pm$0.32\% EER, a 28\% relative reduction over Xiao and Vu~\cite{xiao2025layerwise}'s all-layer ensemble trained on the same data. Across five cross-domain datasets, the average EER is 4.81$\pm$0.17\% with only 1.34M trainable parameters. Detailed comparisons with state-of-the-art systems are presented in Section~\ref{sec:sota}.

We make three contributions:
\begin{enumerate}
\item A model-agnostic probing methodology that uses lightweight XGBoost classifiers to rank transformer layers by cross-domain discriminative power and identify informative depth zones \emph{before} training any downstream classifier. To our knowledge, this is the first work to perform principled layer selection as a preprocessing step for audio deepfake detection.
\item Empirical evidence that informative layers cluster in depth \emph{zones} rather than at uniquely optimal positions, and that the probing adapts to each backbone's representational structure, selecting qualitatively different subsets across three architectures.
\item Competitive cross-domain performance with a fraction of the model complexity: four selected layers and 1.34M trainable parameters match or improve upon published systems that use all layers or fine-tune with substantially more parameters.
\end{enumerate}

\section{Related Work}

\subsection{Audio Deepfake Detection}

Research on audio deepfake detection has advanced through the ASVspoof challenge series, and each edition revealed a new failure mode. ASVspoof 2019 LA~\cite{wang2019asvspoof} established a standardized evaluation with 19 TTS and VC attacks recorded in clean conditions. Top-performing systems achieved near-zero EER on this closed set, but these results did not generalize beyond those 19 attacks. ASVspoof 2021~\cite{liu2023asvspoof2021} introduced the Deepfake (DF) partition, which combines ASVspoof 2019 LA material with voice conversion outputs from VCC 2018 and VCC 2020 , all compressed through varying codecs. This required detectors to handle both unseen voice conversion systems and codec-induced acoustic degradation. ASVspoof5~\cite{wang2024asvspoof5} compounded the challenge by introducing crowdsourced recordings from approximately 2,000 speakers under diverse conditions, attacks generated with 32 different algorithms, and, for the first time, adversarial perturbations designed to fool countermeasure systems directly. Across these editions, the consistent finding is that detectors learn cues specific to their training distribution rather than transferable properties of synthetic speech.

Both handcrafted and learned feature extractors share this vulnerability. Two families of techniques have been proposed to address it, each targeting a different source of overfitting. One-class learning methods such as OC-Softmax~\cite{zhang2021oneclass} reformulate the training objective to compact bonafide representations in the embedding space, reducing dependence on knowledge of specific attack algorithms. However, the learned features themselves remain tied to the acoustic conditions of the training corpus. Data augmentation methods such as RawBoost~\cite{tak2022rawboost}, room impulse response convolution~\cite{ko2017study}, and codec simulation help bridge the channel gap between clean training data and noisy real-world recordings by exposing the model to reverberation, compression, and background noise during training. However, augmentation only diversifies acoustic conditions, not synthesis algorithms, so the learned features remain tied to the vocoder signatures present in the training data. Both strategies modify training data or objectives while leaving the underlying feature extractor unchanged. A different direction, explored in the next section, leverages self-supervised speech models that provide richer representations from the start.
\subsection{SSL-Based Detection and Layer Usage Strategies}

Frozen self-supervised speech models generalize better than handcrafted or end-to-end learned features because their representations are not conditioned on any specific set of attacks. Tak et al.~\cite{tak2022wav2vec} were among the first to demonstrate this, replacing handcrafted front-ends with fine-tuned Wav2Vec~2.0 features and outperforming conventional baselines on ASVspoof 2021 LA and DF. Laakkonen et al.'s MLDG-LoRA~\cite{laakkonen2025mldg} adapts an XLSR-53 backbone via LoRA adapters inserted into every attention head, using meta-learning to encourage domain-invariant features across six attack types. However, the adapted representations can only generalize across the attack diversity seen during meta-training, whereas a frozen backbone preserves the full generality learned from hundreds of thousands of hours of diverse speech.

Given a frozen backbone, the remaining question is which of its 24 layers to use. The Introduction outlined three layer-usage strategies — single-layer selection, all-layer fusion, and per-layer classifiers — and their limitations. Here we examine each in more technical detail. Single-layer approaches~\cite{pascu2024,wang2024vocoded} are simple but restrict the detector to one level of abstraction; the more interesting design question is how to combine multiple layers effectively. Among all-layer approaches, Mart\'in-Do\~nas et al.~\cite{martindonas2024} learn a weighted combination of all 25 hidden states via NN-ASP, Pan et al.~\cite{pan2024attentive} apply squeeze-excitation attention to merge all 24 WavLM layer embeddings, and Tran et al.~\cite{tran2025multiconv} aggregate all XLS-R hidden states through SwiGLU gating and feed them into a multi-kernel gated convolution classifier with a CKA dissimilarity loss. All three include every layer in the fusion without distinguishing which contribute useful cross-domain information and which introduce noise. Zhang et al.~\cite{zhang2024sensitive} take a different direction by learning per-layer attention weights via a squeeze-excitation mechanism jointly optimized with the XLS-R backbone. While this allows the model to emphasize certain layers, the selection is performed \emph{inside} the trained model rather than as a preprocessing step, so it cannot inform architecture design before training begins. Xiao and Vu~\cite{xiao2025layerwise} train 25 separate one-class softmax classifiers and sum the output scores at inference. Their post-hoc analysis — using learned attention weights, consecutive center similarity, and per-layer EER — reveals that layers 4--8 and 18--21 are most important while layers 22--24 are the least informative, but the architecture still uses all 25 classifiers with equal score contribution rather than removing those their own analysis identifies as uninformative.

\subsection{Layer Analysis and Probing in Self-Supervised Models}

Probing classifiers are lightweight models trained on frozen representations to diagnose what each layer encodes. First introduced in NLP~\cite{conneau2018probing,tenney2019bert,belinkov2022probing}, they were adapted to speech by Pasad et al.~\cite{pasad2021layerwise,pasad2023comparative}, who found a consistent layer-wise gradient across Wav2Vec~2.0, HuBERT, and related models: early layers encode acoustic and spectral properties, while late layers encode linguistic content. Since synthesis artifacts span these depths, from vocoder traces in early layers to prosodic and linguistic irregularities in late layers, layer selection directly controls which types of fakes the detector can recognize.

Serrano et al.~\cite{serrano2025layer} perform a systematic layer-wise analysis across six frozen SSL backbones on four out-of-domain evaluation corpora. Their findings show that intermediate layers consistently outperform deeper layers across all six models. For XLS-R, the best single layer (layer 6) achieves 5.6\% EER on In-The-Wild, while all-layer MHFA pooling degrades to 9.8\%. However, their selection relies on manual inspection of per-layer EER curves, which requires per-backbone analysis and lacks a generalizable criterion.

El~Kheir et al.~\cite{elkheir2025layerwise} conduct an extensive layer-wise analysis across six SSL models, ten datasets in three languages, and multiple deepfake types. With frozen backbones, they train classifiers using a learned weighted sum across all layers and observe that layers~4--7 receive the highest learned weights for large models. In separate reduced-layer experiments, they show that truncating the backbone to the first 10--12 layers matches or exceeds full-model performance, concluding that lower layers are most important for deepfake detection. However, their layer importance is derived from weights learned during in-domain training, not from evaluating which layers contribute most to cross-domain generalization. Sinha et al.~\cite{sinha2024layerconditioned} similarly found that layer~16 of HuBERT Large yielded the best discriminative performance, further reinforcing that intermediate layers encode the most task-relevant information.

As noted in the Introduction, existing layer analyses — whether manual, learned in-domain, or post-hoc~\cite{xiao2025layerwise} — reveal layer importance only after the full model is trained. We address this by using XGBoost probing classifiers as a preprocessing step to identify the most discriminative layers before training the downstream classifier. Table~\ref{tab:layer_comparison} summarizes the architectural differences across layer usage strategies.

\begin{table}[!htb]
\centering
\caption{Comparison of layer utilization strategies for SSL-based deepfake detection.}
\label{tab:layer_comparison}
\footnotesize
\setlength{\tabcolsep}{3pt}
\begin{tabular}{lccc}
\toprule
\textbf{Method} & \textbf{Layers} & \textbf{Selection} & \textbf{Fusion} \\
\midrule
Single-layer~\cite{pascu2024} & 1 & Manual & --- \\
NN-ASP~\cite{martindonas2024} & 25 & None & Feature (wt.) \\
Att.\ merge~\cite{pan2024attentive} & 24 & None & Feature (att.) \\
Xiao \& Vu~\cite{xiao2025layerwise} & 25 & Post-hoc & Decision (sum) \\
\makecell[l]{Serrano et al.\\~\cite{serrano2025layer}} & 1 / all & \makecell{Oracle\\/ MHFA} & \makecell{None\\/ Feature} \\
El~Kheir et al.~\cite{elkheir2025layerwise} & all & Learned (e2e) & Feature (wt.) \\
Tran et al.~\cite{tran2025multiconv} & 25 & None (gated) & Feature (gated) \\
\textbf{Ours} & \textbf{4} & \textbf{Probing} & \textbf{Feature (concat)} \\
\bottomrule
\end{tabular}
\end{table}

\section{Proposed Method}
\label{sec:method}

\begin{figure*}[!t]
\centering
\includegraphics[width=\textwidth]{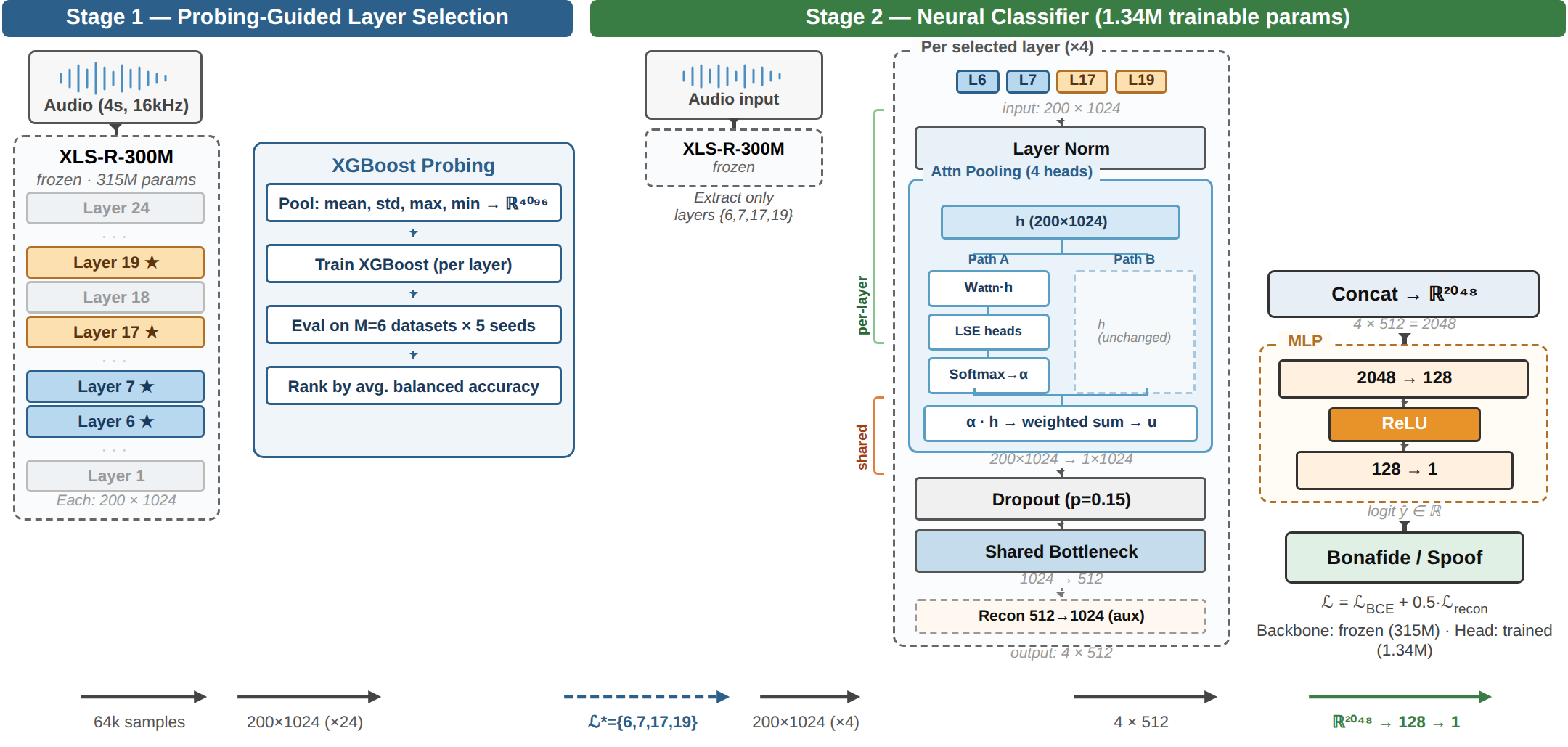}
\caption{Overview of the proposed two-stage approach, illustrated with XLS-R-300M as the backbone. \textbf{Stage~1} (left): the frozen XLS-R-300M backbone extracts hidden representations from all 24 transformer layers; lightweight XGBoost probes, trained on ASVspoof 2019 LA, independently evaluate each layer on $M{=}6$ evaluation datasets, and the top-performing subset $\mathcal{L}^*{=}\{6,\,7,\,17,\,19\}$ is selected via balanced-accuracy ranking. \textbf{Stage~2} (right): each selected layer's hidden states pass through per-layer normalization and four-head attention pooling, then dropout and a shared bottleneck projection ($1024{\to}512$). The four 512-dimensional embeddings are concatenated into a 2048-dimensional vector and classified by a two-layer MLP ($2048{\to}128{\to}1$). A dashed reconstruction branch ($512{\to}1024$) provides auxiliary regularization. The backbone remains completely frozen; only the 1.34M-parameter classifier head is trained.}
\label{fig:pipeline}
\end{figure*}

Our approach consists of two stages, illustrated in Fig.~\ref{fig:pipeline}. In Stage~1, we extract hidden states from all $L$ transformer layers of a frozen SSL backbone, then evaluate each layer using lightweight XGBoost probes~\cite{chen2016xgboost}, rank layers by their cross-domain discriminative power, and finally select a compact subset. In Stage~2, we extract only the selected layers' hidden states, process each through attention pooling and a shared bottleneck projection, concatenate the resulting embeddings, and train a single neural classifier.

\subsection{Stage 1: Probing-Guided Layer Selection}
\label{sec:stage1}

We require a probe that (i)~evaluates each layer independently, (ii)~measures cross-domain generalization rather than in-domain accuracy, and (iii)~is lightweight enough that performance differences reflect layer quality, not classifier capacity.

\subsubsection{Probing Setup}
Given a frozen SSL backbone with $L$ transformer layers, we extract hidden states from all layers for every utterance in the ASVspoof 2019 LA training set (the CNN feature extractor output is excluded, as it encodes low-level acoustic features~\cite{pasad2021layerwise} with limited cross-domain discriminative value~\cite{xiao2025layerwise}). For layer~$l$, the hidden-state sequence $\mathbf{H}^{(l)}_i = [\mathbf{h}^{(l)}_{i,1}, \dots, \mathbf{h}^{(l)}_{i,T}] \in \mathbb{R}^{T \times d}$ is reduced to a single utterance-level representation by concatenating four temporal statistics:
\begin{multline}
    \mathbf{x}_i^{(l)} = \bigl[\text{mean}_t(\mathbf{H}^{(l)}_i);\; \text{std}_t(\mathbf{H}^{(l)}_i);\; \\
    \max_t(\mathbf{H}^{(l)}_i);\; \min_t(\mathbf{H}^{(l)}_i)\bigr] \in \mathbb{R}^{4d},
\label{eq:meanpool}
\end{multline}
where each statistic is computed element-wise over the time dimension, yielding a $d$-dimensional vector, and the four vectors are concatenated.

We then train an XGBoost classifier~\cite{chen2016xgboost} on these pooled representations with binary labels $y_i \in \{0, 1\}$ (spoof vs.\ bonafide). We use 200 estimators with maximum depth~6, learning rate~0.1, and 80\% subsampling of both rows and features. XGBoost is well suited for this role because it is model-agnostic (requiring no architecture design), and expressive enough to capture nonlinear structure in the pooled statistics (std, max, min) without the capacity confound introduced by neural classifiers. We use XGBoost rather than a linear probe because preliminary experiments showed logistic regression achieved near-chance accuracy on cross-domain datasets, confirming the need for a nonlinear probe. To ensure stability, each layer--dataset combination is evaluated across $S{=}5$ random seeds.

\subsubsection{Cross-Domain Evaluation}
Each probe is evaluated on $M{=}6$ test sets: five cross-domain datasets (In-The-Wild~\cite{muller2022itw}, ASVspoof 2021 DF~\cite{yamagishi2021asvspoof}, FakeAVCeleb~\cite{khalid2021fakeavceleb}, WaveFake~\cite{frank2021wavefake}, and ASVspoof5~\cite{wang2024asvspoof5}) plus the in-family ASVspoof 2019 LA Eval partition. All six datasets contribute equally to the composite score. The ASVspoof 2019 LA Dev partition is excluded because it serves as the validation set for early stopping in Stage~2. The composite score for layer~$l$ is the average balanced accuracy across all seeds and datasets:
\begin{equation}
    \text{Score}(l) = \frac{1}{S \cdot M} \sum_{s=1}^{S} \sum_{m=1}^{M} \balacc_m^{(l,s)}.
\label{eq:score}
\end{equation}
Algorithm~\ref{alg:probing} summarizes the full procedure.

\begin{algorithm}[t]
\caption{Probing-Guided Layer Selection}
\label{alg:probing}
\begin{algorithmic}[1]
\STATE \textbf{Input:} Frozen SSL backbone, training set $\mathcal{D}_{\text{train}}$, test datasets $\{\mathcal{D}_1, \dots, \mathcal{D}_M\}$, seeds $\{s_1, \dots, s_S\}$
\STATE \textbf{Output:} Selected layer set $\mathcal{L}^*$
\FOR{each layer $l \in \{1, 2, \dots, L\}$}
    \STATE Extract $\mathbf{H}^{(l)} = \text{SSL}_l(\mathcal{D}_{\text{train}})$
    \STATE Pool: $\mathbf{x}_i^{(l)} = [\text{mean}; \text{std}; \max; \min]_t(\mathbf{H}^{(l)}_i) \in \mathbb{R}^{4d}$
    \FOR{each seed $s \in \{s_1, \dots, s_S\}$}
        \STATE Train XGBoost on $\{(\mathbf{x}_i^{(l)}, y_i)\}$
        \FOR{each test dataset $\mathcal{D}_m$}
            \STATE Compute $\balacc_m^{(l,s)}$
        \ENDFOR
    \ENDFOR
    \STATE $\text{Score}(l) = \frac{1}{S \cdot M} \sum_{s,m} \balacc_m^{(l,s)}$
\ENDFOR
\STATE Rank layers by $\text{Score}(l)$ in descending order
\STATE Select top-$K$ layers where marginal gain plateaus
\RETURN $\mathcal{L}^*$
\end{algorithmic}
\end{algorithm}

\subsubsection{Layer Ranking and Selection}
The probing results yield a layer ranking sorted by $\text{Score}(l)$. We select $K$ layers based on two converging signals: a cluster boundary in the probing scores where marginal gains plateau, and independent confirmation from the Stage~2 ablation that adding lower-ranked layers does not improve CD-Avg. The concrete instantiation of this selection for three backbones is reported in Section~\ref{sec:probing_results}.

\subsection{Stage 2: Neural Classifier}
\label{sec:stage2}

Given the selected layer set $\mathcal{L}^*$, Stage~2 trains a compact neural classifier that fuses features from the selected layers at the representation level. The full architecture is shown in Fig.~\ref{fig:pipeline} (Stage~2).

\subsubsection{Feature Extraction}
For each input utterance, we forward the waveform through the frozen SSL backbone and extract the hidden states corresponding to the selected layers. Each layer~$l \in \mathcal{L}^*$ produces a sequence $\mathbf{H}^{(l)} \in \mathbb{R}^{T \times d}$, where $T$ is the number of frames (determined by the input length and the model's downsampling factor of 320 samples per frame at 16\,kHz).

\subsubsection{Per-Layer Processing}
Each layer's hidden-state sequence passes through layer normalization and multi-head attention pooling (both with per-layer weights), followed by a shared bottleneck projection.

\paragraph{Layer Normalization} We apply layer normalization~\cite{ba2016layernorm} independently to each layer's hidden states. This is necessary because different transformer layers operate at different scales, and normalization ensures that the downstream attention mechanism treats all layers comparably.

\paragraph{Multi-Head Attention Pooling} To aggregate the variable-length frame sequence into a fixed-dimensional utterance embedding, we use a multi-head attention pooling mechanism with $N_h{=}4$ heads. For layer~$l$, a linear projection $\mathbf{W}^{(l)}_{\text{attn}} \in \mathbb{R}^{d \times N_h}$ maps each frame to $N_h$ attention scores. These scores are aggregated across heads via log-sum-exp, then normalized with softmax over the time dimension:
\begin{equation}
    \alpha_t^{(l)} = \frac{\exp\!\bigl(\text{LSE}_h(\mathbf{h}_t^{(l)\top} \mathbf{W}^{(l)}_{\text{attn}})\bigr)}{\sum_{t'} \exp\!\bigl(\text{LSE}_h(\mathbf{h}_{t'}^{(l)\top} \mathbf{W}^{(l)}_{\text{attn}})\bigr)},
\label{eq:attn}
\end{equation}
where $\text{LSE}_h(\cdot) = \log\sum_j \exp([\cdot]_j)$ aggregates across the $N_h$ head dimensions. The utterance-level embedding for layer~$l$ is the weighted sum $\mathbf{u}^{(l)} = \sum_t \alpha_t^{(l)} \mathbf{h}_t^{(l)} \in \mathbb{R}^{d}$.

\paragraph{Bottleneck Projection} Dropout ($p{=}0.15$) is applied to the pooled embedding before a shared linear projection maps it to a 512-dimensional bottleneck:
\begin{equation}
    \mathbf{f}^{(l)} = \mathbf{W}_b \,\text{Dropout}\bigl(\mathbf{u}^{(l)}\bigr) + \mathbf{b}_b \in \mathbb{R}^{512},
\label{eq:bottleneck}
\end{equation}
where $\mathbf{W}_b \in \mathbb{R}^{512 \times d}$ and $\mathbf{b}_b \in \mathbb{R}^{512}$ are shared across all selected layers. Sharing weights forces all layers to project into a common embedding space, which regularizes the model and reduces parameter count. An auxiliary reconstruction head $\mathbf{W}_r \in \mathbb{R}^{d \times 512}$ maps each bottleneck embedding back to $d$ dimensions, and the mean squared error between the reconstruction and the (detached) dropout-masked pooled embedding provides a regularization signal against over-compression to a classification-only subspace.

\subsubsection{Feature-Level Fusion and Classification}
The four bottleneck embeddings are concatenated into a single vector $\mathbf{z} = [\mathbf{f}^{(l_1)}; \dots; \mathbf{f}^{(l_K)}] \in \mathbb{R}^{K \cdot 512}$. A two-layer MLP maps this vector to a scalar logit:
\begin{equation}
    \hat{y} = \mathbf{w}_2^\top \,\text{ReLU}(\mathbf{W}_1 \mathbf{z} + \mathbf{b}_1) + b_2,
\label{eq:mlp}
\end{equation}
with $\mathbf{W}_1 \in \mathbb{R}^{128 \times (K \cdot 512)}$ and $\mathbf{w}_2 \in \mathbb{R}^{128}$. The total training objective combines binary cross-entropy on the logit with the reconstruction loss:
\begin{equation}
    \mathcal{L} = \mathcal{L}_{\text{BCE}}(\hat{y}, y) + \lambda \cdot \mathcal{L}_{\text{recon}},
\label{eq:loss}
\end{equation}
where $\mathcal{L}_{\text{recon}} = \frac{1}{|\mathcal{L}^*|}\sum_{l \in \mathcal{L}^*} \text{MSE}(\mathbf{W}_r \mathbf{f}^{(l)},\; \text{sg}(\tilde{\mathbf{u}}^{(l)}))$, $\tilde{\mathbf{u}}^{(l)} = \text{Dropout}(\mathbf{u}^{(l)})$ is the dropout-masked pooled embedding, $\text{sg}(\cdot)$ denotes stop-gradient, and $\lambda{=}0.5$.

The classifier parameter count scales linearly with $K$ (e.g., 1.34M for $K{=}4$ with $d$-dimensional hidden states). The SSL backbone remains completely frozen throughout training, so total inference cost is dominated by a single forward pass through the backbone, the same cost incurred by any system that uses all $L$ layers, since hidden states are computed incrementally.

\section{Experimental Setup}
\label{sec:setup}

\subsection{Datasets}

Table~\ref{tab:datasets} summarizes all datasets used in this work. We train exclusively on the ASVspoof 2019 LA training partition~\cite{wang2019asvspoof}, which contains 25,380 utterances (2,580 bonafide, 22,800 spoofed) generated by 6 TTS and 13 VC systems. All models are evaluated on seven test sets spanning three categories.

\emph{In-domain evaluation.} The ASVspoof 2019 LA Dev and Eval partitions share the same recording conditions and attack families as the training set. Dev serves as a validation checkpoint; Eval introduces held-out attack algorithms but remains in-family. We report their EERs separately and exclude them from the cross-domain average.

\emph{Cross-domain evaluation.} Five datasets assess generalization to unseen sources, codecs, and attack types (see Table~\ref{tab:datasets} for details). The cross-domain average EER is computed over these five datasets. We additionally report results on the ASVspoof5 Dev partition, which shares the same data and attack algorithms as ASVspoof5 Eval but lacks adversarial perturbations, providing a controlled contrast to isolate the effect of adversarial attacks.

\begin{table}[!htb]
\centering
\caption{Dataset summary.}
\label{tab:datasets}
\small
\setlength{\tabcolsep}{2.5pt}
\begin{tabular}{lcrrc}
\toprule
\textbf{Dataset} & \textbf{Role} & \textbf{Bona.} & \textbf{Spoof} & \makecell{\textbf{Attack}\\[-1pt]\textbf{types}} \\
\midrule
\makecell[l]{ASVspoof 2019\\LA Train} & Train & 2,580 & 22,800 & \makecell{6 TTS\\+ 13 VC} \\
\midrule
\makecell[l]{ASVspoof 2019\\LA Dev} & \makecell{In-\\family} & 2,548 & 22,296 & same \\[2pt]
\makecell[l]{ASVspoof 2019\\LA Eval} & \makecell{In-\\family} & 7,355 & 63,882 & held-out \\
\midrule
In-The-Wild & \makecell{Cross-\\domain} & 19,963 & 11,816 & \makecell{mixed\\(web)} \\[2pt]
ASVspoof 2021 DF & \makecell{Cross-\\domain} & 22,617 & 589,212 & \makecell{TTS+VC\\(codec)} \\[2pt]
FakeAVCeleb & \makecell{Cross-\\domain} & 10,209 & 11,335 & TTS+VC \\[2pt]
WaveFake & \makecell{Cross-\\domain} & 13,100 & 117,983 & \makecell{6\\vocoders} \\[2pt]
ASVspoof5 Eval & \makecell{Cross-\\domain} & 138,688 & 542,086 & \makecell{diverse +\\adversarial} \\[2pt]
ASVspoof5 Dev & Ref.$^\dagger$ & 31,334 & 109,616 & \makecell{diverse\\(no adv.)} \\
\bottomrule
\multicolumn{5}{l}{\scriptsize $^\dagger$Dev vs.\ Eval contrast; not included in cross-domain average.}
\end{tabular}
\end{table}

\subsection{Data Augmentation}
\label{sec:augmentation}

To improve cross-domain generalization, we apply a sequence of augmentation operations during training. Before augmentation, leading and trailing silence is trimmed following Xiao and Vu's~\cite{xiao2025layerwise} finding that silence regions can carry dataset-specific artifacts that hurt cross-domain generalization. We then apply three strategies sequentially: room impulse response (RIR) convolution with probability $1/3$, using real room impulse responses from the OpenSLR-28 corpus~\cite{ko2017study} to simulate reverberant environments; random cropping to 4-second segments, with zero-padding for shorter utterances; and additive Gaussian noise ($\sigma{=}0.005$) with probability $1/5$, providing mild perturbations that discourage the classifier from relying on fine-grained spectral details. No augmentation is applied during inference.

\subsection{Training Details}
\label{sec:training}

The XLS-R-300M backbone~\cite{babu2022xlsr}\footnote{\tt facebook/wav2vec2-xls-r-300m} remains fully frozen throughout training. All three backbones evaluated in this work (XLS-R-300M, WavLM Large, XLSR-53) share a 24-layer transformer architecture with hidden dimension $d{=}1024$. Only the classifier head (1.34M parameters) is updated. We use the Adam optimizer with a learning rate of $5 \times 10^{-5}$ and weight decay $10^{-4}$. The learning rate is halved every 5 epochs via a step scheduler (StepLR, $\gamma{=}0.5$). Training runs for a maximum of 15 epochs with early stopping based on validation EER (patience~5 epochs, monitored on ASVspoof 2019 LA Dev). The batch size is 32, and class-balanced sampling rebalances the class distribution at each epoch to counteract the ${\sim}9{:}1$ spoof-to-bonafide ratio in the training set. Gradients are clipped to a maximum norm of 1.0. All experiments are repeated with three random seeds $\{42, 123, 456\}$, and we report mean $\pm$ standard deviation.

\subsection{Baselines}
\label{sec:baselines}

We compare against three systems (Table~\ref{tab:comparison}), all trained on ASVspoof 2019 LA. Xiao and Vu~\cite{xiao2025layerwise} serve as the primary baseline because they use the same backbone (XLS-R-300M) and comparable augmentation, differing primarily in layer usage strategy (all 25 layers with decision-level fusion vs.\ our 4 selected layers with feature-level fusion). MLDG-LoRA~\cite{laakkonen2025mldg} provides contrast as a backbone-adaptation approach (LoRA + meta-learning, 3.59M parameters). Tran et al.~\cite{tran2025multiconv} test whether full fine-tuning with all-layer gating (318M parameters) outperforms frozen-backbone layer selection. Architectural details for each system are described in Section~2.2.

\subsection{Computational Cost}
\label{sec:cost}

The Stage~1 probing completes in approximately one hour on CPU using pre-extracted features (5 seeds $\times$ 24 layers $\times$ 6 evaluation datasets). Because the XGBoost probes operate on pooled statistics ($\mathbb{R}^{4d}$ per utterance) rather than raw hidden-state sequences, training and evaluation are fast; the additional cost is a one-time feature extraction pass through the frozen backbone to cache all 24 layers' representations, which is performed once and reused across all probing experiments.

\section{Results}
\label{sec:results}

\subsection{XLS-R-300M Layer Probing Results}
\label{sec:probing_results}

Applied to XLS-R-300M, the probing yields a clear ranking. The top four layers by cross-domain balanced accuracy are: layer~19 (79.0\%), layer~6 (77.7\%), layer~17 (76.5\%), and layer~7 (76.4\%), followed by a plateau beginning at layer~11 (76.2\%). This cluster boundary provides a natural cutoff at $K{=}4$, independently confirmed by the ablation in Section~\ref{sec:ablation}. The selected set $\mathcal{L}^* = \{6, 7, 17, 19\}$ spans both mid-depth representations (layers~6 and~7) and late-depth representations (layers~17 and~19), and aligns with the layer importance patterns Xiao and Vu~\cite{xiao2025layerwise} identified post-hoc.

\begin{figure}[!htb]
\centering
\includegraphics[width=\columnwidth]{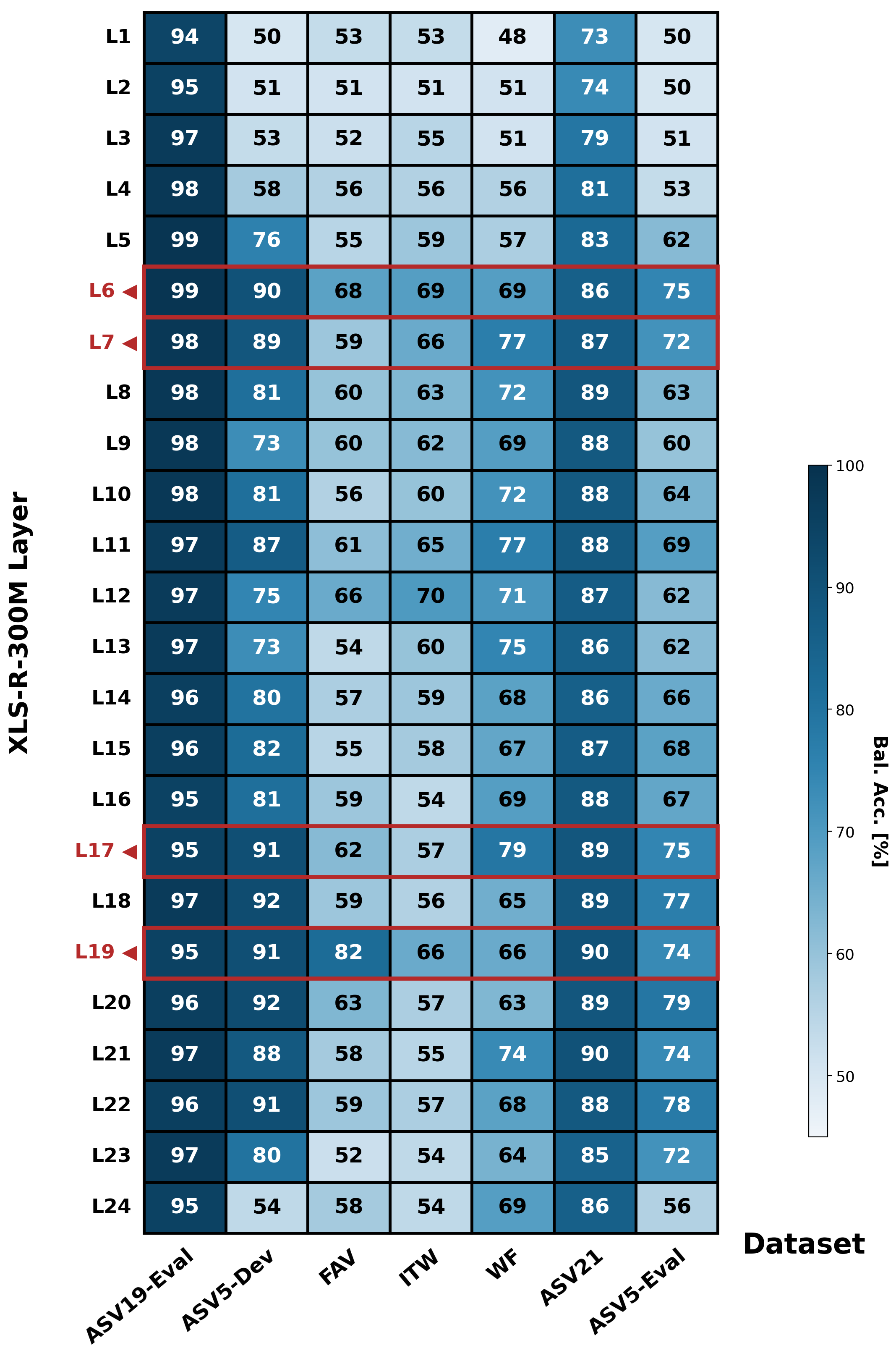}
\caption{Balanced accuracy per layer $\times$ dataset (mean over 5 seeds).
No single layer dominates all domains, motivating multi-layer fusion.
Selected layers $\{6,7,17,19\}$ are marked with $\blacktriangleleft$.
ASVspoof5 Dev and Eval are shown separately. The pronounced accuracy
drop in the Eval column reflects adversarial perturbations present
only in that partition.}
\label{fig:layer_heatmap}
\end{figure}

The heatmap (Fig.~\ref{fig:layer_heatmap}) reveals the key pattern underlying these scores: no single layer dominates across all evaluation sets. Different layers lead on different datasets, confirming that the selected layers capture complementary artifact types and motivating feature-level fusion rather than reliance on any single layer. The deepest layers (23--24) perform worst, consistent with Xiao and Vu's finding that these layers produce EER above 30\% on In-The-Wild.

To assess whether the layer selection is driven by any single evaluation dataset, we performed a leave-one-dataset-out stability analysis on the probing ranking. Layers~6 and~19 appeared in the top four across all six folds, and layer~17 in four of six, confirming their robustness. Layer~7 appeared in two folds, with layer~11, which occupies the same mid-depth representational zone, as the most common substitute. In all cases where one of the four selected layers dropped from the top four, the margin was less than 1 percentage point, and the ablation study (Table~\ref{tab:ablation}) independently confirms that swapping L7 for L11 changes CD-Avg by only 0.10~pp (4.59\% $\to$ 4.69\%). The selection is thus stable at the level of representational depth zones (mid-depth + late-depth), with minor variation in which specific mid-depth layer is included.

\subsection{Cross-Backbone Probing Results}
\label{sec:cross_backbone_probing}

To validate that the probing methodology generalizes beyond XLS-R-300M, we applied the same Stage~1 procedure to two additional frozen backbones: WavLM Large~\cite{chen2022wavlm} (pretrained on 94k hours of primarily English speech with a denoising objective) and XLSR-53 (pretrained on 56k hours of speech across 53 languages with a contrastive objective). Both share the same 24-layer architecture as XLS-R-300M. All probing hyperparameters, evaluation datasets, and scoring criteria were held constant.

Table~\ref{tab:cross_backbone_probing} reports the top-4 layer selections and resulting zone patterns for each backbone.

\begin{table}[!htb]
\centering
\caption{Probing-guided layer selection across three backbones. All backbones share the same 24-layer, 1024-dim architecture but differ in pretraining data and objectives. The same Stage~1 procedure is applied to each; only the selected layers change.}
\label{tab:cross_backbone_probing}
\footnotesize
\setlength{\tabcolsep}{2pt}
\begin{tabular}{llcl}
\toprule
\textbf{Backbone} & \textbf{Pretraining} & \textbf{Top-4} & \textbf{Zone} \\
\midrule
XLS-R-300M & \makecell[l]{436k hrs, 128 langs\\contrastive} & $\{6,7,17,19\}$ & Mid + Late \\[4pt]
WavLM Large & \makecell[l]{94k hrs, English\\contrastive + denoising} & $\{7,8,9,10\}$ & Mid only \\[4pt]
XLSR-53 & \makecell[l]{56k hrs, 53 langs\\contrastive} & $\{14,17,20,21\}$ & Late only \\
\bottomrule
\end{tabular}
\end{table}

Three observations emerge. First, the probing selects qualitatively different layer subsets on each backbone. XLS-R-300M uses two zones (mid-depth layers~6--7 and late-depth layers~17--19), WavLM Large concentrates entirely in a single mid-depth zone (layers~7--10), and XLSR-53 shifts toward late-depth layers (14, 17, 20, 21). This shows that the probing produces backbone-specific selections rather than a fixed layer recipe.

Second, the zone differences correlate with pretraining differences: WavLM Large's late-layer probing scores plateau at 69--70\% compared to 77--78\% for mid-depth layers, while XLS-R-300M shows strong scores at both mid and late depths. Whether these patterns are driven by pretraining objectives, data scale, or other factors remains an open question.

Third, the probing indicates that transferring XLS-R-300M's layer indices to other backbones would be suboptimal. Layers~17 and~19, which rank \#3 and \#1 on XLS-R-300M, rank \#16 and \#19 on WavLM Large (probing scores of 70.1\% and 69.2\%, respectively). A practitioner who skipped the probing step and reused $\{6,7,17,19\}$ on WavLM would include two layers from the weakest zone of that backbone.

Together, these results indicate that per-backbone probing is a necessary step rather than an optional refinement: the informative zones differ across architectures, and reusing one backbone's selection on another risks including uninformative layers.

\begin{table*}[!t]
\centering
\caption{Equal Error Rate (EER, \%) across all evaluation datasets for three random seeds. The cross-domain average is computed over the five cross-domain datasets (excluding in-domain, in-family, and reference partitions). Best mean per dataset is in \textbf{bold}.}
\label{tab:main_results}
\small
\begin{tabular}{llcccl}
\toprule
\textbf{Category} & \textbf{Dataset} & \textbf{Seed 42} & \textbf{Seed 123} & \textbf{Seed 456} & \textbf{Mean$\pm$Std} \\
\midrule
In-domain & ASVspoof 2019 LA Dev & 0.13 & 0.12 & 0.12 & 0.12$\pm$0.01 \\
In-family & ASVspoof 2019 LA Eval & 3.41 & 3.29 & 2.96 & 3.22$\pm$0.19 \\
\midrule
\multirow{5}{*}{\makecell[l]{Cross-\\domain}}
& In-The-Wild & 4.87 & 4.58 & 5.37 & 4.94$\pm$0.32 \\
& ASVspoof 2021 DF & 3.24 & 3.13 & 3.32 & 3.23$\pm$0.07 \\
& FakeAVCeleb & 0.97 & 0.94 & 1.32 & 1.08$\pm$0.17 \\
& WaveFake & 3.91 & 3.52 & 3.86 & 3.76$\pm$0.17 \\
& ASVspoof5 Eval & 11.20 & 10.75 & 11.14 & 11.03$\pm$0.20 \\
\midrule
Reference$^\dagger$ & ASVspoof5 Dev & 0.90 & 0.82 & 0.87 & 0.86$\pm$0.04 \\
\midrule
& \textbf{Cross-domain average} & 4.84 & 4.59 & 5.00 & \textbf{4.81$\pm$0.17} \\
\bottomrule
\multicolumn{6}{l}{\scriptsize $^\dagger$ASVspoof5 Dev is reported for reference (Dev vs.\ Eval contrast) and is not included in the cross-domain average.}
\end{tabular}
\end{table*}

\subsection{Cross-Domain Detection Performance}
\label{sec:main_results}

Table~\ref{tab:main_results} reports EER across all evaluation datasets for three random seeds. The cross-domain average is \textbf{4.81$\pm$0.17\%}, with low variance across seeds (0.17~pp), confirming stability across initializations.

Performance varies substantially across datasets, ranging from near-solved (FakeAVCeleb, 1.08\%) to challenging (ASVspoof5 Eval, 11.03\%). The elevated ASVspoof5 Eval EER reflects adversarial perturbations designed to evade countermeasure systems~\cite{wang2024asvspoof5}. The contrast with ASVspoof5 Dev is striking: sharing the same data and attacks but lacking adversarial perturbations, it achieves just 0.86$\pm$0.04\% EER, confirming that this gap is driven by adversarial attacks rather than domain shift. ASVspoof5 Dev was not used in any stage of the pipeline, so this result is fully independent.

The in-family ASVspoof 2019 LA Eval achieves 3.22$\pm$0.19\%, reflecting the expected trade-off of frozen-backbone approaches discussed in Section~\ref{sec:sota}.
\subsection{Comparison with State-of-the-Art}
\label{sec:sota}

Table~\ref{tab:comparison} compares our system against four baselines, all trained on ASVspoof 2019 LA. Since each system evaluates on a different subset of benchmarks, cross-domain averages are computed over the four shared datasets (ITW, ASV21 DF, FakeAVCeleb, ASVspoof5) to ensure a fair comparison. Three patterns emerge.

First, probing-guided layer selection with feature-level fusion improves over exhaustive decision-level fusion under matched conditions. Against Xiao and Vu~\cite{xiao2025layerwise}, who use the same backbone and comparable augmentation but fuse all 25 layers, our four-layer system achieves a 28\% relative EER reduction on In-The-Wild (4.94\% vs.\ 6.90\%).

Second, a frozen backbone trades in-family performance for cross-domain robustness. Our system achieves lower EER than MLDG-LoRA~\cite{laakkonen2025mldg} on four of five shared cross-domain datasets, but MLDG-LoRA leads on ASVspoof 2019 LA Eval (0.54\% vs.\ 3.22\%), where backbone adaptation to the training distribution provides a direct advantage.

Third, Tran et al.~\cite{tran2025multiconv} achieve lower EER on the three benchmarks they report, with the largest gap on ASV21 DF (1.53\% vs 3.23\%) and a narrow margin on ITW (0.16~pp). Their system, however, fine-tunes 318M parameters compared to our 1.34M, and does not evaluate on FakeAVCeleb, WaveFake, or ASVspoof5.

Ultimately, the choice between these strategies depends on the deployment scenario: in-distribution accuracy favors backbone adaptation, while robustness to unseen attacks favors frozen-backbone layer selection.

\begin{table}[!htb]
\centering
\caption{Architecture and performance comparison with state-of-the-art systems. All methods train on ASVspoof 2019 LA. EER (\%) reported; best per dataset in \textbf{bold}. ``---'' = not evaluated.}
\label{tab:comparison}
\footnotesize
\setlength{\tabcolsep}{2pt}
\begin{tabular}{lcccc}
\toprule
& \makecell{\textbf{Xiao \&}\\[-1pt]\textbf{Vu}~\cite{xiao2025layerwise}} & \makecell{\textbf{MLDG-}\\[-1pt]\textbf{LoRA}~\cite{laakkonen2025mldg}} & \makecell{\textbf{Tran}\\[-1pt]\textbf{et al.}~\cite{tran2025multiconv}} & \textbf{Ours} \\
\midrule
\multicolumn{5}{l}{\textit{Architecture}} \\
\midrule
Backbone & XLS-R & W2V 2.0 & XLS-R & XLS-R \\
Backbone status & Frozen & LoRA & Fine-tuned & Frozen \\
Layers used & 25 & All & 25 (gated) & \textbf{4} \\
Trainable params & ${\sim}$25$\times$cls & 3.59M & 318M$^\S$ & \textbf{1.34M} \\
Fusion & Decision (sum) & Feature & Gated & Concat \\
\midrule
\multicolumn{5}{l}{\textit{Performance (EER \%)}} \\
\midrule
ASV Eval & 5.27$\pm$.39 & 0.54$\pm$.33 & \textbf{0.10}$^\ddagger$ & 3.22$\pm$.19 \\[2pt]
ITW & 6.90$\pm$.30 & 6.81$\pm$.81 & \textbf{4.78}$^\ddagger$ & 4.94$\pm$.32 \\[2pt]
ASV21 DF & --- & 3.99$\pm$.46 & \textbf{1.53}$^\ddagger$ & 3.23$\pm$.07 \\[2pt]
FAV & --- & 1.48$\pm$1.08 & --- & \textbf{1.08}$\pm$.17 \\[2pt]
WaveFake & --- & --- & --- & 3.76$\pm$.17 \\[2pt]
ASV5 & --- & 14.10$\pm$.39 & --- & \textbf{11.03}$\pm$.20 \\
\midrule
Shared CD-Avg & --- & 6.60$^*$ & --- & \textbf{5.07}$^*$ \\
\bottomrule
\multicolumn{5}{l}{\scriptsize $^*$Over 4 shared datasets (ITW, ASV21 DF, FAV, ASV5).} \\
\multicolumn{5}{l}{\scriptsize $^\ddagger$Mean of 3 runs. $^\S$Full fine-tuned.}
\end{tabular}
\end{table}

\begin{table*}[!t]
\centering
\caption{Ablation study. All models use seed~123 and the same architecture unless noted. EER~(\%) reported. $\Delta$\% = relative change in CD-Avg vs.\ Proposed.}
\label{tab:ablation}
\footnotesize
\setlength{\tabcolsep}{3pt}
\begin{tabular}{clcccccccccc}
\toprule
\textbf{Quality} & \textbf{Config} & \textbf{Layers} & \textbf{ASV Dev} & \textbf{ASV Eval} & \textbf{ITW} & \textbf{ASV21 DF} & \textbf{FakeAVCeleb} & \textbf{WaveFake} & \textbf{ASV5} & \textbf{CD-Avg} & \textbf{$\Delta$\%} \\
\midrule
\multicolumn{12}{l}{\textit{(a) Worst-case configurations --- probing-rank constraints violated}} \\
\midrule
\cellcolor{gray!15} \checkmark & \cellcolor{gray!15} Proposed & \cellcolor{gray!15} $\layerset$ & \cellcolor{gray!15} 0.12 & \cellcolor{gray!15} 3.29 & \cellcolor{gray!15} 4.58 & \cellcolor{gray!15} 3.13 & \cellcolor{gray!15} 0.94 & \cellcolor{gray!15} 3.52 & \cellcolor{gray!15} 10.75 & \cellcolor{gray!15} \textbf{4.59} & \cellcolor{gray!15} --- \\
\cellcolor{lightred} $\times$ & \cellcolor{lightred} Worst-A (early+late) & \cellcolor{lightred} {[1,2,23,24]}   & \cellcolor{lightred} 0.39 & \cellcolor{lightred} 11.04 & \cellcolor{lightred} 16.56 & \cellcolor{lightred} 20.80 & \cellcolor{lightred} 25.36 & \cellcolor{lightred} 37.36 & \cellcolor{lightred} 26.32 & \cellcolor{lightred} 25.28 & \cellcolor{lightred} +451 \\
\cellcolor{lightred} $\times$ & \cellcolor{lightred} Worst-B (all late)       & \cellcolor{lightred} {[21,22,23,24]} & \cellcolor{lightred} 0.30 & \cellcolor{lightred} 3.20  & \cellcolor{lightred} 13.04 & \cellcolor{lightred} 4.16  & \cellcolor{lightred} 4.44  & \cellcolor{lightred} 10.28 & \cellcolor{lightred} 11.76 & \cellcolor{lightred} 8.74  & \cellcolor{lightred} +90 \\
\cellcolor{lightred} $\times$ & \cellcolor{lightred} Worst-C (all early)      & \cellcolor{lightred} {[1,2,3,4]}     & \cellcolor{lightred} 0.21 & \cellcolor{lightred} 8.80  & \cellcolor{lightred} 8.08  & \cellcolor{lightred} 9.68  & \cellcolor{lightred} 12.28 & \cellcolor{lightred} 11.08 & \cellcolor{lightred} 21.64 & \cellcolor{lightred} 12.55 & \cellcolor{lightred} +173 \\
\cellcolor{lightred} $\times$ & \cellcolor{lightred} Worst-D (outlier)  & \cellcolor{lightred} {[1,2,21,22]}   & \cellcolor{lightred} 0.20 & \cellcolor{lightred} 3.80  & \cellcolor{lightred} 7.68  & \cellcolor{lightred} 6.12  & \cellcolor{lightred} 8.92  & \cellcolor{lightred} 12.92 & \cellcolor{lightred} 14.52 & \cellcolor{lightred} 10.03 & \cellcolor{lightred} +119 \\
\midrule
\multicolumn{12}{l}{\textit{(b) Alternative probing-guided configurations --- within-framework variants}} \\
\midrule
\cellcolor{gray!15} \checkmark & \cellcolor{gray!15} Proposed & \cellcolor{gray!15} $\layerset$ & \cellcolor{gray!15} 0.12 & \cellcolor{gray!15} 3.29 & \cellcolor{gray!15} 4.58 & \cellcolor{gray!15} 3.13 & \cellcolor{gray!15} 0.94 & \cellcolor{gray!15} 3.52 & \cellcolor{gray!15} 10.75 & \cellcolor{gray!15} \textbf{4.59} & \cellcolor{gray!15} --- \\
$\sim$ & Swap L17$\to$L11               & {[6,7,11,19]}    & 0.10 & 3.42 & 4.28 & 2.96 & 1.16 & 4.48 & 10.56 & 4.69 & +2.2 \\
$\sim$ & Swap L17$\to$L8                & {[6,7,8,19]}     & 0.13 & 3.44 & 4.88 & 3.52 & 1.52 & 3.68 & 10.40 & 4.80 & +4.6 \\
$\sim$ & Swap L6$\to$L8                 & {[8,7,17,19]}    & 0.12 & 3.72 & 5.72 & 3.36 & 1.64 & 4.60 & 10.44 & 5.15 & +12.2 \\
$\sim$ & Proposed + L11                 & {[6,7,11,17,19]} & 0.11 & 3.47 & 4.92 & 3.04 & 1.00 & 3.60 & 11.44 & 4.80 & +4.6 \\
$\sim$ & Proposed + L12                 & {[6,7,12,17,19]} & 0.12 & 3.51 & 4.88 & 3.00 & 1.00 & 3.56 & 11.52 & 4.79 & +4.4 \\
$\sim$ & Proposed + L8                  & {[6,7,8,17,19]}  & 0.12 & 3.41 & 5.08 & 3.00 & 1.04 & 3.64 & 11.52 & 4.86 & +5.9 \\
$\sim$ & Top-6                          & {[6,7,11,12,17,19]} & 0.11 & 3.52 & 4.92 & 2.88 & 1.00 & 3.72 & 11.16 & 4.74 & +3.3 \\
\midrule
\multicolumn{12}{l}{\textit{(c) Layer selection strategy comparisons}} \\
\midrule
\cellcolor{gray!15} \checkmark & \cellcolor{gray!15} Proposed (BalAcc) & \cellcolor{gray!15} $\layerset$ & \cellcolor{gray!15} 0.12 & \cellcolor{gray!15} 3.29 & \cellcolor{gray!15} 4.58 & \cellcolor{gray!15} 3.13 & \cellcolor{gray!15} 0.94 & \cellcolor{gray!15} 3.52 & \cellcolor{gray!15} 10.75 & \cellcolor{gray!15} \textbf{4.59} & \cellcolor{gray!15} --- \\
$\times$ & EER-ranked top-4              & {[7,8,9,10]}     & 0.12 & 2.76 & 6.68 & 2.96 & 1.92 & 2.92 & 11.16 & 5.13 & +11.8 \\
$\times$ & Mid-early consec.              & {[4,5,6,7]}      & 0.10 & 4.04 & 4.76 & 3.36 & 2.36 & 5.20 & 13.72 & 5.88 & +28.1 \\
$\times$ & Mid-depth consec.             & {[13,14,15,16]}  & 0.18 & 3.36 & 8.12 & 4.32 & 1.00 & 5.76 & 10.24 & 5.89 & +28.3 \\
$\times$ & El~Kheir et al.               & {[1,2,3,4,5,6]}  & 0.13 & 5.36 & 4.16 & 4.36 & 3.72 & 6.24 & 14.04 & 6.50 & +41.6 \\
\midrule
\multicolumn{12}{l}{\textit{(d) Architecture ablations}} \\
\midrule
\cellcolor{gray!15} \checkmark & \cellcolor{gray!15} Proposed ($\lambda{=}0.5$, att.) & \cellcolor{gray!15} $\layerset$ & \cellcolor{gray!15} 0.12 & \cellcolor{gray!15} 3.29 & \cellcolor{gray!15} 4.58 & \cellcolor{gray!15} 3.13 & \cellcolor{gray!15} 0.94 & \cellcolor{gray!15} 3.52 & \cellcolor{gray!15} 10.75 & \cellcolor{gray!15} \textbf{4.59} & \cellcolor{gray!15} --- \\
$\sim$ & $\lambda{=}0$ (no recon loss)  & $\layerset$      & 0.10 & 3.92 & 4.64 & 3.08 & 1.24 & 4.64 & 11.36 & 4.99 & +8.7 \\
$\times$ & Mean pooling                  & $\layerset$      & 0.20 & 3.20 & 5.48 & 4.28 & 1.84 & 4.36 & 11.68 & 5.53 & +20.5 \\
\midrule
\multicolumn{12}{l}{\textit{(e) Layer count scaling}} \\
\midrule
$\times$ & Single layer                   & {[19]}           & 0.22 & 3.08 & 7.56 & 4.48 & 3.40 & 9.48 & 9.80 & 6.94 & +51.2 \\
$\sim$ & Two layers                     & {[6,19]}         & 0.13 & 3.13 & 4.24 & 2.99 & 1.07 & 2.82 & 11.51 & 4.53 & $-$1.3 \\
$\sim$ & Three layers                   & {[6,7,19]}       & 0.11 & 3.28 & 4.16 & 3.76 & 1.20 & 5.12 & 10.32 & 4.91 & +7.0 \\
\cellcolor{gray!15} \checkmark & \cellcolor{gray!15} Proposed (4 layers) & \cellcolor{gray!15} $\layerset$ & \cellcolor{gray!15} 0.12 & \cellcolor{gray!15} 3.29 & \cellcolor{gray!15} 4.58 & \cellcolor{gray!15} 3.13 & \cellcolor{gray!15} 0.94 & \cellcolor{gray!15} 3.52 & \cellcolor{gray!15} 10.75 & \cellcolor{gray!15} \textbf{4.59} & \cellcolor{gray!15} --- \\
\bottomrule
\multicolumn{12}{l}{\checkmark~optimal; $\sim$~suboptimal; $\times$~poor. $\Delta$\% = relative change in CD-Avg vs.\ Proposed (4.59\%).}
\end{tabular}
\end{table*}

\subsection{Methodology Transfer Across Backbones}
\label{sec:methodology_transfer}

To confirm that the probing-guided pipeline transfers to other backbones, we trained the same Stage~2 classifier on WavLM Large and XLSR-53 using each backbone's probing-selected layers (Table~\ref{tab:transfer_results}).  Worst-case layer selections degrade CD-Avg by 1.8--5.5$\times$ across all three backbones (Table~\ref{tab:cross_backbone_worst}), showing that the benefit of probing-guided selection holds regardless of backbone.

\begin{table}[!htb]
\centering
\caption{Cross-backbone comparison using probing-selected layers. CD-Avg = mean EER across five cross-domain datasets (seed~123). All systems use identical Stage~2 architecture and augmentation; only the backbone and selected layers differ.}
\label{tab:transfer_results}
\footnotesize
\setlength{\tabcolsep}{3pt}
\begin{tabular}{llccc}
\toprule
\textbf{Backbone} & \textbf{Layers} & \textbf{Params} & \textbf{CD-Avg} & \textbf{ITW} \\
\midrule
XLS-R-300M & $\{6,7,17,19\}$ & 1.34M & \textbf{4.59\%} & 4.58\% \\
WavLM Large & $\{7,8,9,10\}$ & 1.34M & 7.14\% & 9.68\% \\
XLSR-53 & $\{14,17,20,21\}$ & 1.34M & 8.91\% & 13.01\% \\
\bottomrule
\end{tabular}
\end{table}

\begin{table}[!htb]
\centering
\caption{Cross-backbone worst-case analysis. CD-Avg (\%) for probing-selected vs worst-case layer configurations (seed~123).}
\label{tab:cross_backbone_worst}
\scriptsize
\setlength{\tabcolsep}{2pt}
\begin{tabular}{@{}llccc@{}}
\toprule
\textbf{Config} & \textbf{Layers} & \textbf{XLS-R} & \textbf{WavLM} & \textbf{XLSR-53} \\
\midrule
Probing & --- & 4.59 & 7.14 & 8.91 \\
Worst-A & $\{1,2,23,24\}$ & 25.28\,(5.5$\times$) & 12.66\,(1.8$\times$) & 31.80\,(3.6$\times$) \\
Worst-B & $\{21,22,23,24\}$\textsuperscript{\dag} & 8.74\,(1.9$\times$) & 12.83\,(1.8$\times$) & 19.41\,(2.2$\times$) \\
Worst-C & $\{1,2,3,4\}$ & 12.55\,(2.7$\times$) & 16.31\,(2.3$\times$) & 16.10\,(1.8$\times$) \\
\bottomrule
\multicolumn{5}{@{}l@{}}{\textsuperscript{\dag}XLSR-53 uses $\{3,22,23,24\}$ because L21 is a probing-selected layer.}
\end{tabular}
\end{table}

\subsection{Ablation Studies}
\label{sec:ablation}

\begin{figure*}[!t]
\centering
\includegraphics[width=\textwidth]{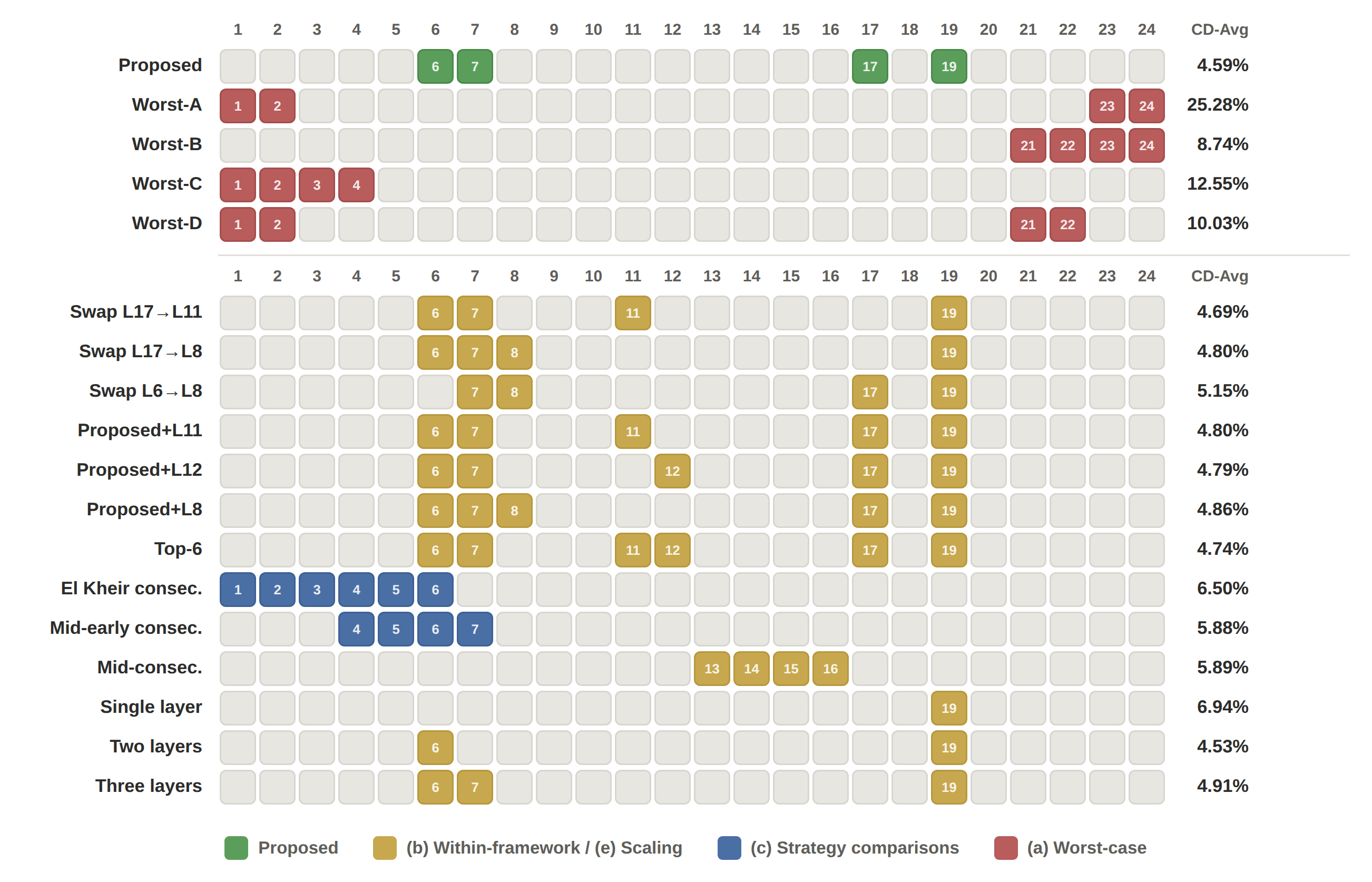}
\caption{Spatial view of layer selections across all ablation configurations.
Each row shows which of the 24 XLS-R layers are selected (colored) or not (gray).
\textbf{(a)}~Worst-case configurations cluster at the extremes of the network (very early or
very late layers), avoiding the mid-depth region identified by probing as most informative.
\textbf{(b)}~All probing-guided alternatives concentrate in the same mid- and late-depth zones
as the proposed $\layerset$, confirming the robustness of the probing signal.
CD-Avg EER (\%) shown on the right.}
\label{fig:layer_span}
\end{figure*}

\begin{figure}[!t]
\centering
\includegraphics[width=\columnwidth]{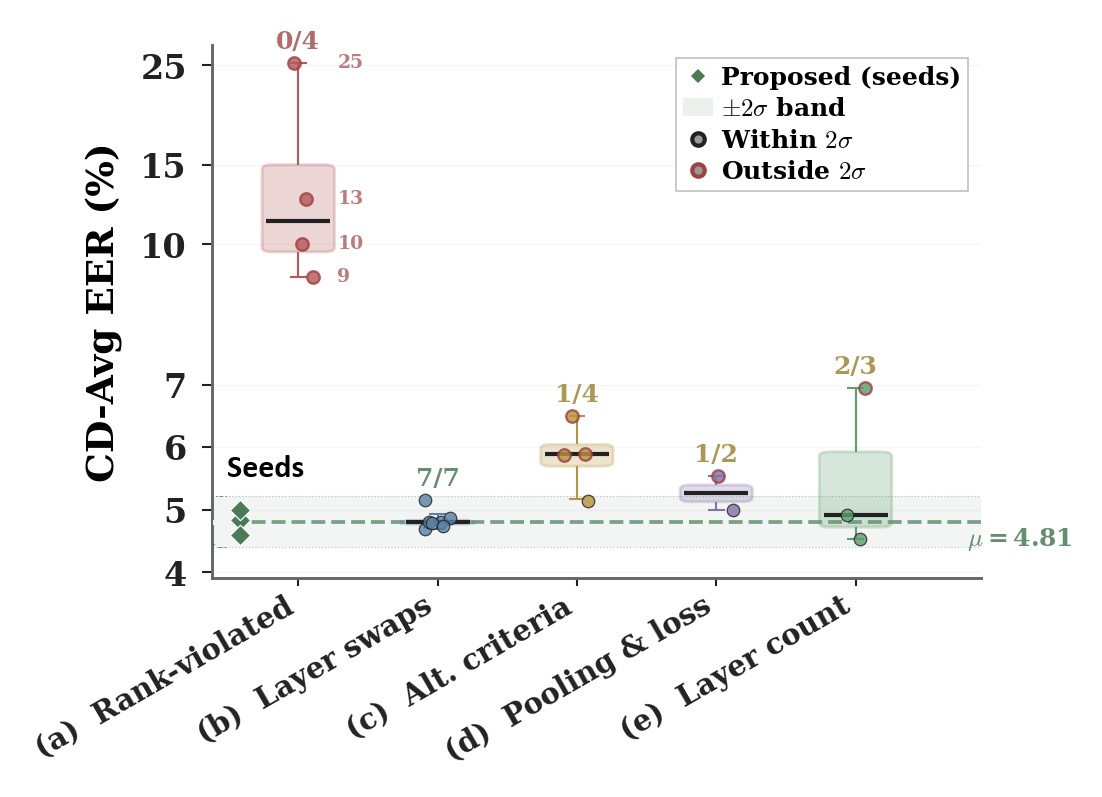}
\caption{CD-Avg EER across ablation categories. The green band shows the $\pm 2\sigma$ range from the proposed system's three-seed evaluation (diamonds, left margin). Circles indicate configurations within 2$\sigma$; squares indicate those outside. All seven within-framework layer swaps (b) fall within seed noise, while rank-violated (a) and alternative-criteria (c) configurations fall clearly outside.}
\label{fig:ablation_boxplot}
\end{figure}

The ablation study addresses a central question: does the probing identify \emph{specific layers} that are uniquely optimal, or does it identify \emph{depth zones} from which any reasonable combination performs comparably? To answer this, we compare the proposed $\layerset$ against 20 alternative configurations spanning five categories (Table~\ref{tab:ablation}). All experiments use seed~123 and are otherwise identical. Fig.~\ref{fig:layer_span} provides a spatial view of which layers are selected in each case, and Fig.~\ref{fig:ablation_boxplot} summarizes CD-Avg across all ablation categories relative to the proposed system's three-seed variance ($\pm 2\sigma$ band: 4.47--5.15\%).

\paragraph{Layer selection matters: the probing-guided set outperforms alternatives}
The initial probing selected layers $\layerset$, spanning mid-depth (L6, L7) and late-depth (L17, L19) representations. This configuration achieves 4.59\% CD-Avg (seed~123). Table~\ref{tab:ablation}(a) confirms that violating the probing ranking leads to large degradations: all four worst-case configurations achieve very low ASVspoof Dev EER (0.20\%--0.39\%), yet their cross-domain performance collapses to 8.74--25.28\% CD-Avg ($\Delta$\%: +90\% to +451\%). Worst-A, for instance, achieves 0.39\% Dev EER yet 25.28\% CD-Avg---the lowest in-domain error paired with the worst cross-domain performance of any configuration tested. This confirms that very early and very late layers lack transferable discriminative information. Therefore, the probing step is necessary to distinguish transferable from non-transferable layers before classifier training. Table~\ref{tab:ablation}(c) further shows that alternative selection criteria---consecutive blocks or applying El~Kheir et al.'s first-$N$ layer selection strategy~\cite{elkheir2025layerwise} to our architecture---degrade CD-Avg by 28--42\%.

\paragraph{Within-zone robustness: depth zones matter, not exact layers}

A clear pattern emerges from the within-framework ablations. Table~\ref{tab:ablation}(b) tests seven variants that swap, add, or remove layers while remaining within the mid- and late-depth zones where the probing-selected layers cluster. As Fig.~\ref{fig:ablation_boxplot} shows, all seven fall within the $\pm 2\sigma$ noise band of the proposed system's multi-seed variance (7/7 within 2$\sigma$). The relative changes range from just +2.2\% to +12.2\%---comparable to seed-to-seed variation. The EER-ranked selection $\{7,8,9,10\}$ from Table~\ref{tab:ablation}(c) further supports this: it produces comparable CD-Avg (5.13\%) while concentrating in a single depth zone. Together, these results confirm that informative layers cluster in depth \emph{zones} rather than at uniquely optimal positions.

Several observations support this interpretation. First, the two-layer configuration [6,19]---one mid-depth, one late-depth---already achieves 4.53\% CD-Avg, within the noise band of the proposed four-layer system, suggesting that a single representative from each zone captures most of the cross-domain signal. Second, adding layer~11 to the proposed set ($K{=}5$) does not improve CD-Avg (4.80\%), because L11 occupies the same mid-depth zone as L6 and L7 and introduces redundant rather than complementary information. Third, the three-layer configuration [6,7,19] yields 4.91\%---adding a second mid-depth layer without a corresponding late-depth layer shifts the balance toward one zone without introducing new representational content. The four-layer proposed set achieves 4.59\% by adding L17, providing representation from both zones.

\paragraph{Architecture ablations}
Table~\ref{tab:ablation}(d) isolates two architectural choices. Removing the reconstruction loss ($\lambda{=}0$) increases CD-Avg by 0.40~pp (4.99\%), with the largest effect on WaveFake (4.64\% vs.\ 3.52\%), indicating that the auxiliary signal preserves information that would otherwise be lost in bottleneck compression. Replacing attention pooling with mean pooling increases CD-Avg by 0.94~pp (5.53\%), suggesting that learned frame-level weighting matters for identifying discriminative temporal regions. 

\section{Discussion}
\label{sec:discussion}

\paragraph{Layer selection identifies depth zones, not uniquely optimal layers} The central finding is that probing identifies informative \emph{depth zones} rather than uniquely optimal layers. All within-zone substitutions fall within seed noise, while zone violations degrade performance by up to 5$\times$ (Section~\ref{sec:ablation}). The practical implication is that a practitioner applying this methodology to a new backbone need not find the exact top-$K$ layers; identifying the correct depth zones and sampling one or two representatives from each is sufficient. This also explains why uninformative layers hurt rather than merely contribute nothing: in decision-level ensembles where each layer contributes equally, the deepest layers---which encode pretraining-specific abstractions rather than transferable artifact cues---dilute the signal from informative zones.

\paragraph{Frozen backbone and cross-backbone generalizability} Keeping the backbone frozen trades in-family performance for cross-domain robustness (Section~\ref{sec:sota}). When the attacks at deployment are unpredictable, a frozen backbone offers more robust generalization. When the attack types are known in advance, backbone adaptation can exploit that knowledge for better in-distribution accuracy. A hybrid strategy---applying lightweight adaptation (e.g., LoRA) to the selected layers only---could close the in-family gap without sacrificing cross-domain robustness. The probing methodology itself generalizes across backbones: applied to three architectures with different pretraining objectives and data scales, it selects qualitatively different zone patterns (Section~\ref{sec:cross_backbone_probing}), confirming that the methodology adapts to each backbone's representational structure rather than producing a fixed recipe.

\paragraph{Limitations} Our probing uses ASVspoof 2019 LA as training data and evaluates on five cross-domain datasets. The optimal layer selection may shift as new attack types emerge. While the top-four selection is straightforward for XLS-R, automating subset size selection through greedy forward selection or mutual-information-based criteria would further generalize the pipeline. Robustness to adversarial attacks remains a limitation, as perturbations in ASVspoof5 Eval are optimized to corrupt the layer representations our model relies on, and no layer selection strategy alone can fully address this. Adversarial training or input purification are natural directions for future work. One potential concern is that the probing ranking uses the same evaluation datasets as the final system. However, the XGBoost probes and the neural classifier share no parameters or training signals, and the ablation results provide independent confirmation of the $K{=}4$ choice. ASVspoof5 Dev, which was not included in the probing evaluation, achieves 0.86$\pm$0.04\% EER with the selected layers, and the leave-one-dataset-out analysis (Section~\ref{sec:stage1}) shows that layers~6 and~19 remain in the top four regardless of which dataset is excluded.

\section{Conclusion}
\label{sec:conclusion}

We presented a model-agnostic methodology for identifying which depth zones of a frozen self-supervised speech model carry the most transferable representations for audio deepfake detection. The approach uses lightweight XGBoost probes to rank each transformer layer by cross-domain discriminative power \emph{before} any task-specific classifier is trained, then builds a compact neural classifier that operates only on the selected layers.

Applied to XLS-R-300M, the probing selects layers that cluster in mid-depth and late-depth zones, yielding layers $\{6, 7, 17, 19\}$. A neural classifier fusing these four layers achieves 4.94$\pm$0.32\% EER on In-The-Wild and 4.81$\pm$0.17\% cross-domain average. With only 1.34M trainable parameters, it outperforms Xiao and Vu's 25-layer decision-level ensemble and matches Tran et al.'s fine-tuned system. Ablation studies reveal that the selected layers cluster in informative depth \emph{zones} rather than at uniquely optimal positions: all seven within-zone layer substitutions fall within the multi-seed noise band, while zone violations degrade performance by up to 5$\times$. The critical design decision is \emph{which zones} to include, not which exact layers---a finding that directly benefits practitioners applying the methodology to new backbones.

Because the probing stage is independent of the downstream classifier architecture, the methodology generalizes to any multi-layer self-supervised backbone. Validation on WavLM Large and XLSR-53 confirms that the probing adapts to each backbone's representational structure, selecting qualitatively different depth zones (mid-only for WavLM, late-heavy for XLSR-53). Future work will investigate automated layer subset selection and evaluate robustness to emerging neural codec-based attacks.

\section*{Acknowledgements}

\section*{Declaration of competing interest}
The authors declare that they have no known competing financial interests or personal relationships that could have appeared to influence the work reported in this paper.

\section*{Funding}
This research did not receive any specific grant from funding agencies in the public, commercial, or not-for-profit sectors.

\section*{CRediT authorship contribution statement}
\textbf{Marjan Beheshti:} Conceptualization, Methodology, Software, Formal analysis, Investigation, Visualization, Writing -- original draft.
\textbf{Majid Rostami:} Conceptualization, Writing -- review \& editing.
\textbf{Bo Chen:} Supervision, Project administration, Writing -- review \& editing.

\section*{Data availability}
All datasets used in this work are publicly available. ASVspoof 2019 LA, ASVspoof 2021 DF, and ASVspoof5 are available through the ASVspoof consortium. In-The-Wild, FakeAVCeleb, and WaveFake are available from their respective authors. Code and trained models will be made available upon publication.

\section*{Declaration of generative AI and AI-assisted technologies in the manuscript preparation process}
During the preparation of this work the author(s) used Claude (Anthropic) in order to assist with language editing and manuscript formatting. After using this tool, the author(s) reviewed and edited the content as needed and take(s) full responsibility for the content of the published article.

\FloatBarrier

\end{document}